# Fullerene Faraday Cage Keeps Magnetic Properties of Inner Cluster Pristine

Stanislav M. Avdoshenko,*

Correspondence to: Stanislav M. Avdoshenko (E-mail: *s.avdoshenko@gmail.com*)

*Leibniz Institute for Solid State and Materials Research Dresden, 01069 Dresden, Germany,*

**ABSTRACT**

Any single molecular magnets (SMM) perspective for application is as good as its magnetization stability in ambient conditions. Endohedral metallofullerenes (EMFs) provide a solid basis for promising SMMs. In this study, we investigated the behavior of functionalized EMFs on a gold surface (EMF-L-Au). Having followed the systems molecular dynamics paths, we observed that the chemically locked inner cluster inside fullerene cage will remain locked even at room temperature due to the ligand-effect. We have located multiple possible minima with different charge arrangements between EMF-L-Au fragments. Remarkably, the charge state of the EMF inner cluster remained virtually constant and so magnetic properties are expected to be untouched.

## Introduction

Single molecular magnet materials are perspective for various application in spintronics ranging from high-density information storages to dynamic spin-logic elements.[1]–[3] The magnetization stability is a limiting factor in the research, design and use. Internal relaxation channels are important and have to be accounted for. In a real device, however, the ambient can significantly reduce the magnetization stability.[1],[4] Currently, the surface impact on properties of these systems is under theoretical and experimental scrutiny.[5]–[7] Endohedral metallofullerenes provide a solid basis for promising SMMs, where the fullerene cage can play the role of a natural shield from the environment charges.[8]–[10] Timely, in this report, we outline a theoretical prediction for Prato [5,6]-monoadducts of $Me_3N@C_{80}$ behavior on gold surfaces (**EMF-L**-Au for short) using DFT-based molecular dynamics(MD) modeling (NVT dynamics at 300 K and Newton-Raphson quenching of few sampled structures to 0 K). For the 0 K structures, a classical crystal field modeling was made in the attempt to estimate a distribution of magnetic properties in possible ensembles of molecules. The magnetically interesting systems will be the one with $DySc_2N$ or $Dy_2ScN$ clusters inside. The molecular dynamics studies, however, were performed on the $Sc_3N$ containing molecule. Nonetheless, these results can be generalized onto those with $DySc_2N$ or $Dy_2ScN$ due to analogy in chemical behavior. For example, in case of $DySc_2N$ clusters, the *f*-shell does not affect systems dynamics and, as shown early, this leads to a very good agreement between structures observed in theory with *f*-shell free ions substitution and experiments with lanthanides.[8], [11]–[14] To quantify magnetic properties in this work, the structure of $^6H_{15/2}$ multiplets of $Dy^{3+}$ were computed for the quenched structures of **EMF-L**-Au, where each Sc-site was changed to Dy, for the classical crystal field modeling based on Mulliken populations. *The most remarkable result here, that while the quenching of **EMF-L**-Au along MD trajectories lead to very different geometries (local minima) the charges of inner clusters will remain a constant, so will, presumably, the magnetic properties.* Recent experiments conducted for the similar **EMF-L** system on gold surfaces also reported significantly disordered molecular formations on the surfaces. However, even though the quality of formed structures is rather poor, the system is capable of preserving the magnetic properties.

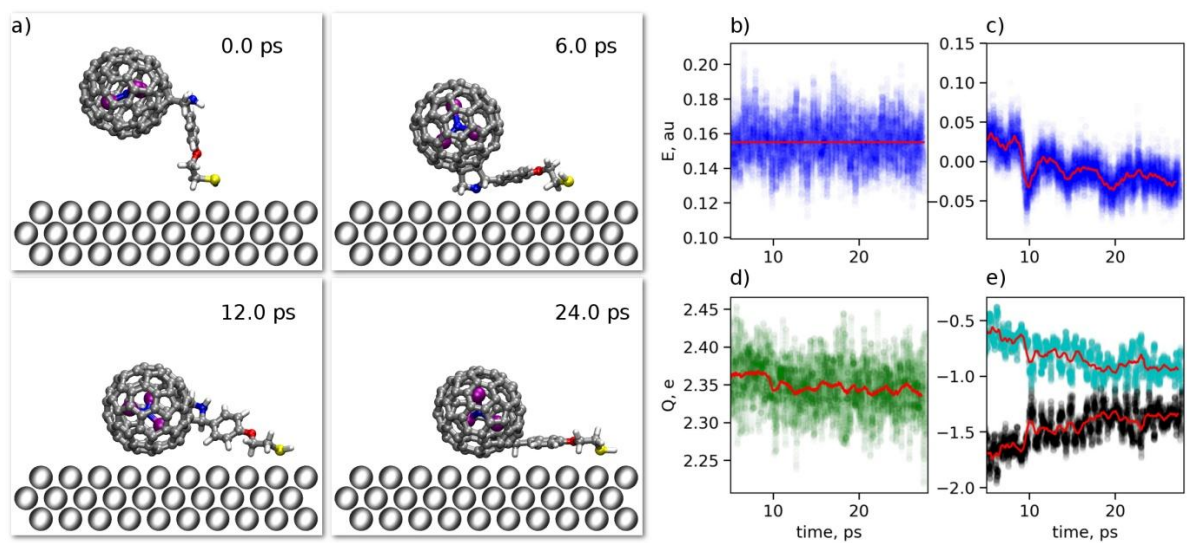

**Figure 1.** a) Shows representative geometries of the EMF-L molecule in close proximity of the gold surface as observed at different times along molecular dynamic trajectory. b) and c) show EMF-L-Au energies change along the MD trajectories kinetic and potential respectively. d) and e) show the time evolution of Mulliken charges in the EMF-L-Au system for different fragments: the metal cluster ($Sc_3N$) charge is plotted in green on d) and the charges of $C_{80}$-L, Au-surface are captured in black and cyan on e), respectively.

## Methods

The L-ligand in question is a Prato product of *N*-methylglycine 2-(4-mercaptoethoxy) benzaldehyde.[15] We have simplified phase space sampling in molecular dynamic studies by changing $CH_3$- in terminal S-$CH_3$ and N-methylpyrrolidine onto H-. The choice of this particular ligand group is driven by its successful use in the design of high quality fullerene-based self-assembled monolayers (SAMs). When applied to an EMF system the behavior on surfaces can change. The EMFs should have different charge arrangements on surfaces, compared to the empty fullerenes, as a result of large differences in chemical potentials they have, which may complicate the SAM production process.

We have designed a geometrical configuration of **EMF-L**-Au as shown in Figure 1a. In these model structures, the gold surface is presented just by 3 atomic layers. Reportedly, the 3-4 layers gold surface is found to be sufficient approximation in the SAMs research.[4] A generation script for the structure with use of Atomistic Simulation Environment libraries for Python is provided.[16] The well optimized **EMF-L** was orthogonal to the gold (this work focuses on FCC(100) surface, but the conclusions are likely to be general for Au(111) or Au(110)) surface and placed such that L-sulfur atom within 0.7 ($R_{Au}^{VdW} + R_S^{VdW}$) distance from the gold surface. During molecular dynamic modeling, as well as the optimization, the inner cluster of the EMF was $Sc_3N$. This is a justifiable approximation due to the known analogy in chemical behavior of Sc/Y and lanthanides.[8], [17]

Born-Oppenheimer molecular dynamics simulation were performed at the PBE-D/DZVP (mesh cutoff 200 Ry) level theory using CP2K code.[18]–[21] The velocity Verlet algorithm was employed with the time step of 0.5 fs and three Nosé-Hoover chain thermostat set at 300 K with the thermostat time constant 100 fs.[22], [23] Although through the molecular dynamics modeling the Au-surface was fixed, in optimization no constrained were applied. In optimized structures the surface reconstruction was insignificant. It leads to a conclusion that MD was not affected by the chosen constraints.

The Siesta code was employed as the optimization driver at PBE/DZP (mesh cutoff 200 Ry) level.[24] The SCF wavefunction was optimized with unrestricted Kohn-Sham method to prove the lack of unintended spin polarization. In-house python scripts were used for analysis and visualization.

Molecular structures, and BOMD trajectories were visualized using VMD package.[25] It is important to note that experimentally thiols (R-S-H) on the Au surfaces are known to dissociate, with covalent bond formation between ligand and surface -R-S-Au.[26]–[28] Theoretical research suggests homolytic cleavage of the S–H bond.[26] Yet, on the reported MD time scale, we were unable to observe spontaneous dissociation process.

Classical (point-charge-based) crystal field calculations were done with McPhase code.[29] In assumed oxidation state of $Dy^{3+}$ the high spin ground state J = 15/2 can be represented in eight low-lying Kramers doublets. For *ab initio* modeling we used Molcas 8.0 code.[30] The active space of the CASSCF calculations includes eleven active electrons and the seven active orbitals (e.g. CAS (11,7)). All 21 sextet states and 108 quartets and only 100 doublets were included in the state-averaged CASSCF procedure and were further mixed by spin-orbit coupling in the RASSI procedure. VDZ quality atomic natural extended relativistic basis set (ANO-RCC) was employed. The single ion magnetic properties and CF parameters were calculated with use of SINGLE ANISO module. The classical and *ab-initio* crystal fields were used to construct a model Zeeman Hamiltonian in $|J,m_J\rangle$-basis (Clebsch-Gordan decomposition). Based on this Hamiltonian transition probabilities were estimated using PHI program.[31]

**The system at 300 K.**

The original system (Figure 1a, and SI) was propagated according to Nosé-Hover equations for 30.0 ps with integration step 1.0 fs, the thermostat temperature was set to 300 K. The systems energy balance during propagation is shown in Figure 1b,1c. Already after the first 5.0 ps, the kinetic energy of the system oscillates around 300 K with a very rare intrusion from the thermostat. The quick descent of the potential energy at that time confirms once again a very unstable nature of vertically aligned **EMF-L** conformation. Our entire attempt to localize a sustained minimum with a vertical **EMF-L** position was in vain. This confirms the fact that EMF-surface interaction is very strong.

However, even after **EMF-L** touchdown, there are still some conformational changes which lead to a significant deviation of the potential energy from a local averages at times (on Figure 1, the red line on all graphs is the running average with time window 1.0 ps). These changes correlate nicely with spikes in charge rearrangements in the system **EMF-L**-Au (Figure 1d, 1e), (usually, around 15.0-20.0 kJ/mol, with the maximum sweep of 30.0 kJ/mol at 10 ps).

Thus, it hardly comes as a surprise, because electron affinities and ionization potentials of both **EMF-L**-Au should be a function of geometric characteristics of **EMF-L**. There is certain regularity in these charge rearrangements, but at this time we did not elucidate any specific extremes with minimum and maximum charge transfer. This will require more in-depth electronic structure analysis and goes beyond the scope of this report.

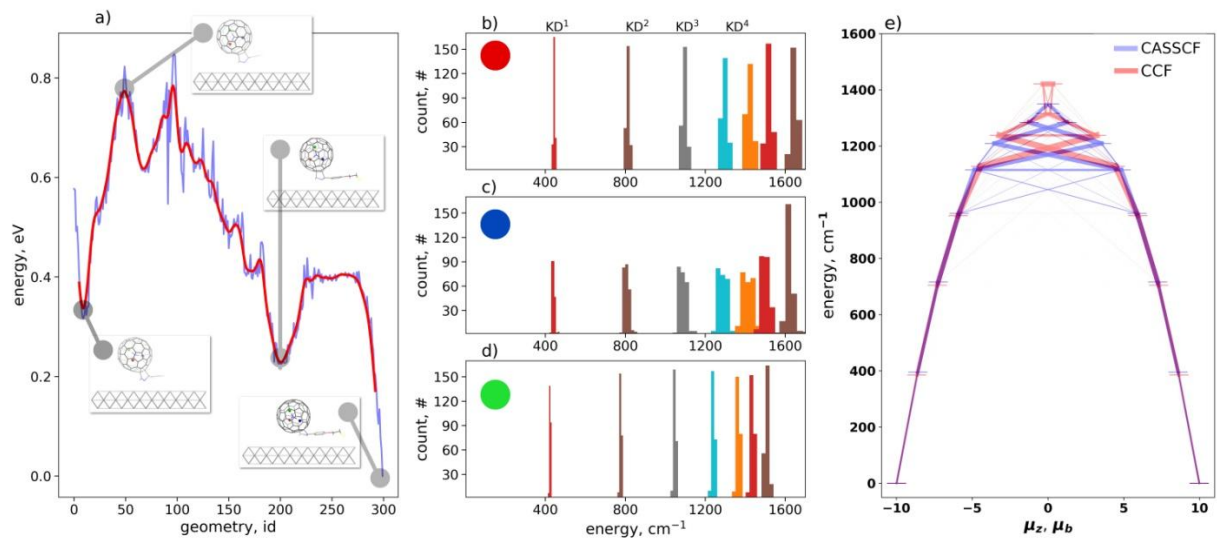

**Figure 2.** a) Potential energy for optimized geometries along MD trajectory, with the sampling rate 100 fs. The light red line is a running average over 10 points. A few energetically notable conformations are rendered b)-d). Histograms of energy distributions for eight Kramers doublets ($KD^1$, $KD^2$, $KD^3$, $KD^4$, *etc.*) constructed based on 300 optimized structures. At each point the KDs energies are estimated classically for $Dy^{3+}$ ion using a point charge crystal field model. Alternating the single $Dy^{3+}$ ion over three different but equally possible positions within $Me_3N$-cluster gives three possible outcomes b)-d). The colored circle in each of the histograms connects them to the color coded positions in the insets on a). e) The of $^6H_{15/2}$ multiplets structures of $Dy^{3+}$ ion as computed *ab initio* at the CASSCF/RASSI level and classical crystal field modeling. The red and blue lines visualize transition probabilities between different $|J,m_J\rangle$ sates.[13],[31]

Thus, at this point, we will just accept the fact that some conformations along the MD path can induce a very quick charge transfer between the **EMF-L** molecule the Au surface and focus on the fact that the charge of the **Me₃N** cluster with **EMF-L** remains almost constant (Figure 1d). This is a very important observation. For example, it suggests right away that magnetization properties in the system should be persevered to a certain extent, as they are mostly defined by the cluster individual properties. In detail, from Figure 1e, one can conclude, that the Au surface leads to a fractional oxidation of the **EMF-L**, with most charge coming from the $C_{80}$ **-L** part. This is natural as **EMF-L** is expected to have a cage-based HOMO orbital, similar to $Me_3N@C_{80}$ molecule.[11] This observation, combined with the fact that in the functionalized $Me_3N@C_{80}$ system the internal cluster is immobilized with the anisotropy axis still well-defined makes such EMF-L based strategy very promising in SMM design.

**The system at 0 K.**

Even if the SAMs quality is uncertain at room temperature, during the magnetization experiments the produced material will be subject to cooling. At this step, any room-temperature state is quenched to experiments temperature (of 1-2 K for XMCD measurements, for example).[8] It might be that during the cooling the ordering can occur like a 2D crystallization. To enable the formation of well-ordered structures at this stage, the individual **EMF-L** building blocks should easily locate a single, well-defined global minimum; otherwise, the following cooling is likely to produce a glass state.

To analyze stability at 0 K we quenched our systems with sampling 100 fs following Newton-Raphson algorithm. We have been able to find a local minimum with an outstanding stability to which all geometries will tend to converge. However, the topology of the potential energy surface is full of local trap-minima, which will complicate the molecular chase for the global minimum. Figure 2a shows the potential energy for the optimized geometries along MD trajectory, with the sampling rate 100 fs. Each point of the blue curve is a well-optimized structure (down to 0.005 eV/Ang) using as a starting point the geometry at the time point t=id*100 fs. Analysis of local maxima and minima along potential energy curve reinforce the conclusion that both parts of **EMF-L** prefer a surface contact energetically (for example, id=300 structure has L and $Me_3N@C_{80}$ in full contact, which is the preferable system state among all considered). Although, each conformation is stable in its own right the rather short 100 fs window between two geometrical points makes the optimized set to be correlated and smooth. However, further running average on 10 id-points was unable to simplify the picture and the potential topology still looks rather complicated. All this indicates a very low likelihood of surface crystallization and clustering during the cooling.

It is important to point out, that the trajectory (Figure 2a) does not just end abruptly, as it may appear. From Figure 1c the system is clearly reaching the saturation, during last 5-10 ps potential energy is hardly changing while geometrically the fullerene starts to bounce on and off the. One has to understand Figure 2a as a set of sampled structures quenched to 0 K where the arrow of time, would be a questionable concept. Moreover, it would be hard to prove if the sampled geometries still can be considered as to be connected to their originator - MD generator.

**Fullerenes Faraday cage**.[9], [10]

Finally, an interesting question to ask at this point is: "how would all this apparent randomness affect the magnetic response of the system"? Possible inside can be given by a simple classical modeling of $Dy^{3+}$ ion multiples in a crystal field of point charges. It was reported a simple scaling of atomic charges can be used to construct a reasonable crystal field which gives eight Kramers doublets (for Dy(III) multiple $^6H_{15/2}$) close to the *ab-initio* solution.[17] Although in more complicated cases classical crystal field can produce wrong results.[32] To make sure that in this system the classical crystal field (CCF) modeling closely reproduces ab initio (CASSCF) result simulated for $DySc_2N@C_{80}$ molecule alone(SI). Figure 2e shows one-to-one comparison of these two $^6H_{15/2}$-multiplet structures. As one can see not only energy but transition probabilities within each multiples have. Here the Mulliken charges were scale by 2.5 to make first Kramers doublet fit the *ab-initio* value. Assuming that to be a general conclusion for any geometrical conformation, we used these 300 well-optimized structures and their atomic Mulliken charges as input for point charge modeling of $Dy^{3+}$ multiples (see Figure 2b-2d). In $[DySc_2N@C_{80}$-L]@gold the $Dy^{3+}$ ion can be placed in three separate locations with respect to the L-group, these sites a captured by the different color on the insets of Figure 2a. The same color "circle" was used on Figure 2b-2d to indicate a respective $Dy^{3+}$ site. Similar to the dynamic picture the charge state of the cluster and the metals remain unperturbed regardless of geometrical or charge arrangements the **EMF-L**@gold has. That fact determines similarity in the multiples spectrum for different sites. Notably, the distribution is very narrow, especially for the lower-lying multiples which are of impotence at the low temperatures of XMCD experiments.[13],[29] Such eventuality leads to a conclusion, that even if the quality of the film is low (dynamic disorder) the magnetic properties will not be impacted or corrupted and will remain uniform along the film.

## Conclusions

At this point, one would wonder "are these observations general or not", "can this protocol be used for different systems"? To answer this question we have applied the same protocol to the **EMF-L**-Au systems with a different ligand group -Ph-S-$CH_3$. It produces the experimentally visible 1,7-$(PhSCH_3)_2$ $DySc_2N@C_{80}$.[33] We have observed virtually identical results (SI, Figure S1). First, the inner cluster trapped inside the cage in one particular orientation. Second, the system preferred geometrical order when both ligand and the fullerene in touch with the gold surface and with point intensive charge rearrangement in the system begins, however, the $Me_3N$ cluster charges stay untouched. Quick screening of magnetic properties distribution using the classical crystal field model on a set of 0K optimized structures will likely lead to the similar results as for the original **EMF-L**-Au. Thus, it makes the outlined theoretical protocol a robust and fast screening tool for EMF-based SMM and maybe other SMM classes too. To fulfill all requirements the sate multiplet of the SMM class in question must be well described in terms of classical crystal.[32]

In brief, Prato [5,6]-monoadducts of $Me_3N@C_{80}$ behavior on gold surface is studied using theoretical protocol which includes DFT-based molecular dynamics and classical crystal field modeling. Due to a narrow distribution of magnetic properties in the system, the report anticipates that EMF-based SMM will produce magnetoresponsive films (surface nanostructures) with properties close to powder experiments, even though films quality is subjected to the ligand chemistry.

## Acknowledgments

I acknowledge funding by the Marie Skłodowska-Curie actions, EU project SamSD (grant agreement number 748635). Computational resources were provided by the centre for Information Services and High Performance Computing (ZIH) in TU Dresden. Also, I thank Dr. Alexey A. Popov for fruitful discussions.

**Keywords:** Molecules on surfaces, single molecular magnets, magnetodynamics, molecular dynamics, density functional theory

Additional Supporting Information may be found in the online version of this article.

## References and Notes

[1] N. Domingo, E. Bellido, D. Ruiz-Molina, *Chem. Soc. Rev.*, **2011**, DOI:10.1039/C1CS15096K.

[2] A. Cornia, M. Mannini, P. Sainctavit, R. Sessoli, *Chem. Soc. Rev.*, **2011**, DOI:10.1039/C0CS00187B.

[3] R. J. Holmberg, M. Murugesu, *J. Mater. Chem. C*, **2015**, DOI:10.1039/C5TC03225C.

[4] A. Lunghi, M. Iannuzzi, R. Sessoli, F. Totti, *J. Mater. Chem. C*, **2015**, DOI:10.1039/C5TC00394F.

[5] M. Mannini, F. Pineider, C. Danieli, F. Totti, L. Sorace, P. Sainctavit, M.-A. Arrio, E. Otero, L. Joly, J. C. Cezar, A. Cornia, R. Sessoli, *Nature*, **2010**, DOI:10.1038/nature09478.

[6] G. F. Garcia, A. Lunghi, F. Totti, R. Sessoli, *Nanoscale*, **2018**, DOI:10.1039/C7NR06320B.

[7] L. Vitali, S. Fabris, A. M. Conte, S. Brink, M. Ruben, S. Baroni, K. Kern, *Nano Lett.*, **2008**, DOI:10.1021/nl801869b.

[8] R. Westerström, J. Dreiser, C. Piamonteze, M. Muntwiler, S. Weyeneth, H. Brune, S. Rusponi, F. Nolting, A. Popov, S. Yang, L. Dunsch, T. Greber, *J. Am. Chem. Soc.*, **2012**, DOI:10.1021/ja301044p.

[9] B. Pietzak, M. Waiblinger, T. A. Murphy, A. Weidinger, M. Höhne, E. Dietel, A. Hirsch, *Chem. Phys. Lett.*, **1997**, DOI:10.1016/S0009-2614(97)01100-7.

[10] P. Delaney, J. C. Greer, *Appl. Phys. Lett.*, **2004**, DOI:10.1063/1.1640783.

[11] B. Elliott, A. D. Pykhova, J. Rivera, C. M. Cardona, L. Dunsch, A. A. Popov, L. Echegoyen, *J. Phys. Chem. C*, **2013**, DOI:10.1021/jp310024u.

[12] R. Valencia, A. Rodríguez-Fortea, A. Clotet, C. de Graaf, M. N. Chaur, L. Echegoyen, J. M. Poblet, *Chem. – Eur. J.*, **2009**, DOI:10.1002/chem.200900728.

[13] D. S. Krylov, F. Liu, S. M. Avdoshenko, L. Spree, B. Weise, A. Waske, A. U. B. Wolter, B. Büchner, A. A. Popov, *Chem. Commun.*, **2017**, DOI:10.1039/C7CC03580B.

[14] S. Yang, A. A. Popov, C. Chen, L. Dunsch, *J. Phys. Chem. C*, **2009**, DOI:10.1021/jp9005263.

[15] M. del C. Gimenez-Lopez, M. T. Räisänen, T. W. Chamberlain, U. Weber, M. Lebedeva, G. A. Rance, G. A. D. Briggs, D. Pettifor, V. Burlakov, M. Buck, A. N. Khlobystov, *Langmuir ACS J. Surf. Colloids*, **2011**, DOI:10.1021/la200654n.

[16] A. H. Larsen, J. J. Mortensen, J. Blomqvist, I. E. Castelli, R. Christensen, Marcin Dułak, J. Friis, M. N. Groves, B. Hammer, C. Hargus, E. D. Hermes, P. C. Jennings, P. B. Jensen, J. Kermode, J. R. Kitchin, E. L. Kolsbjerg, J. Kubal, Kristen Kaasbjerg, S. Lysgaard, J. B. Maronsson, T. Maxson, T. Olsen, L. Pastewka, Andrew Peterson, C. Rostgaard, J. Schiøtz, O. Schütt, M. Strange, K. S. Thygesen, Tejs Vegge, L. Vilhelmsen, M. Walter, Z. Zeng, K. W. Jacobsen, *J. Phys. Condens. Matter*, **2017**, DOI:10.1088/1361-648X/aa680e.

[17] Y. Zhang, D. Krylov, M. Rosenkranz, S. Schiemenz, A. A. Popov, *Chem. Sci.*, **2015**, DOI:10.1039/C5SC00154D.

[18] J. P. Perdew, M. Ernzerhof, K. Burke, *J. Chem. Phys.*, **1996**, DOI:10.1063/1.472933.

[19] S. Grimme, *Wiley Interdiscip. Rev. Comput. Mol. Sci.*, **2011**, DOI:10.1002/wcms.30.

[20] J. VandeVondele, J. Hutter, *J. Chem. Phys.*, **2007**, DOI:10.1063/1.2770708.

[21] J. Hutter, M. Iannuzzi, F. Schiffmann, J. VandeVondele, *Wiley Interdiscip. Rev. Comput. Mol. Sci.*, **2014**, DOI:10.1002/wcms.1159.

[22] S. Nosé, *J. Chem. Phys.*, **1984**, DOI:10.1063/1.447334.

[23] W. G. Hoover, B. L. Holian, *Phys. Lett. A*, **1996**, DOI:10.1016/0375-9601(95)00973-6.

[24] J. M. Soler, E. Artacho, J. D. Gale, A. García, J. Junquera, P. Ordejón, Daniel Sánchez-Portal, *J. Phys. Condens. Matter*, **2002**, DOI:10.1088/0953-8984/14/11/302.

[25] W. Humphrey, A. Dalke, K. Schulten, *J. Mol. Graph.*, **1996**, DOI:10.1016/0263-7855(96)00018-5.

[26] G. Rajaraman, A. Caneschi, D. Gatteschi, F. Totti, *Phys. Chem. Chem. Phys.*, **2011**, DOI:10.1039/C0CP02042G.

[27] H. Häkkinen, *Nat. Chem.*, **2012**, DOI:10.1038/nchem.1352.

[28] E. Pensa, E. Cortés, G. Corthey, P. Carro, C. Vericat, M. H. Fonticelli, G. Benítez, A. A. Rubert, R. C. Salvarezza, *Acc. Chem. Res.*, **2012**, DOI:10.1021/ar200260p.

[29] M. Rotter, *J. Magn. Magn. Mater.*, **2004**, DOI:10.1016/j.jmmm.2003.12.1394.

[30] F. Aquilante, J. Autschbach, R. K. Carlson, L. F. Chibotaru, M. G. Delcey, L. De Vico, I. Fdez. Galván, N. Ferré, L. M. Frutos, L. Gagliardi, M. Garavelli, A. Giussani, C. E. Hoyer, G. Li Manni, H. Lischka, D. Ma, P. Å. Malmqvist, T. Müller, A. Nenov, M. Olivucci, T. B. Pedersen, D. Peng, F. Plasser, B. Pritchard, M. Reiher, I. Rivalta, I. Schapiro, J. Segarra-Martí, M. Stenrup, D. G. Truhlar, L. Ungur, A. Valentini, S. Vancoillie, V. Veryazov, V. P. Vysotskiy, O. Weingart, F. Zapata, R. Lindh, *J. Comput. Chem.*, **2016**, DOI:10.1002/jcc.24221.

[31] N. F. Chilton, R. P. Anderson, L. D. Turner, A. Soncini, K. S. Murray, *J. Comput. Chem.*, **2013**, DOI:10.1002/jcc.23234.

[32] L. Ungur, L. F. Chibotaru, *Chem. – Eur. J.*, **2017**, DOI:10.1002/chem.201605102.

[33] M. Toganoh, K. Suzuki, R. Udagawa, A. Hirai, M. Sawamura, E. Nakamura, *Org. Biomol. Chem.*, **2003**, DOI:10.1039/B302468G.